# Few-cycle optical solitary waves in cascaded-quadratic-cubic-quintic nonlinear media

Kanchan Kumar De[1], Amit Goyal[1], C. N. Kumar[1] and Amarendra K. Sarma[2*]

[1]Department of Physics, Panjab University, Chandigarh-160014, India
[2]Department of Physics, Indian Institute of Technology Guwahati, Guwahati-781 039, Assam, India
*aksarma@iitg.ernet.in

We study the propagation of few-cycle optical solitary waves in a nonlinear media under the combined action of quadratic, cubic and quintic nonlinearities in a large phase-mismatched second harmonic (SHG) process. Exact bright and dark soliton solutions to the nonlinear evolution equation for cascaded quadratic media beyond the slowly varying envelope approximations is reported. The analytical solutions obtained are verified through numerical simulations.

(Some figures may appear in colour only in the online journal)

Considerable research effort is made in the investigation of few optical pulses owing to the recent progress in the generation of few-cycle and even sub-cycle optical pulses [1-6]. Such ultrashort optical pulses have possible applications in many diverse areas such as, high-order harmonic generation, ultrafast spectroscopy, metrology, medical diagnostics and imaging, testing of high-speed devices, optical communications, manipulation of chemical reactions and bond formation, material processing, investigations of light-matter interactions and attosecond physics [1,2]. In this context, it is clear that study of few-cycle pulse propagation phenomena in various media is of tremendous importance. Till recently, most of the pulse propagation models were derived to describe many-cycle pulses using the so called slowly varying envelope approximation (SVEA)[7]. However, it is now well established that the description of few-cycle pulses requires a modification of the SVEA. Many propagation models have been proposed and successfully studied to a varying degree [1-2,4]. Most of these models are aimed at describing few-cycle pulses in cubic nonlinear media[8-13]. However, recently the analysis of ultrashort optical pulse propagation in second-order nonlinear material is getting tremendous boost, particularly media with cascaded-quadratic nonlinearity is drawing particular attention[14-21]. In 2006, Moses and Wise have derived a coupled propagation equations for ultrashort pulses in a degenerate three-wave mixing process in quadratic ($\chi^{(2)}$) media [14]. Moses-Wise model is restricted to the case of strongly mismatched interaction where the conversion efficiency to second or higher harmonics is negligible. In fact, a more generalized nonlinear envelope equation is derived by Conforti et al. to describe the propagation of broadband optical pulses in second order nonlinear materials [15,16]. However, Moses and Wise went on to present, using cascaded quadratic nonlinearity, theoretical and experimental evidence of a new quadratic effect, namely the controllable self-steepening (SS) effect. The controllability of the SS effect is very useful in nonlinear propagation of ultrashort pulses as it may be used to cancel the propagation effects of group velocity mismatch. It is pointed out that the cascaded nonlinearity induces a Raman-like term into the model owing to the nonlocal nature of the cascaded nonlinearity [21]. It may be noted that, traditionally, the intensity dependent refraction (IDR) effects in quadratic media are not expected in quadratically nonlinear media owing to the phase mismatch of the fundamental harmonic with the higher ones within the SVEA [22]. It is possible to impress Kerr-like nonlinear phase-shifts of controllable sign and magnitude on optical pulses with cascaded-quadratic ($\chi^{(2)}:\chi^{(2)}$) nonlinearity [23]. A quadratic medium can be exploited to mimic the propagation of ultrashort pulses under third-order nonlinearity ($\chi^{(3)}$). The propagation obeys an effective nonlinear Schrodinger equation (NLSE) having features not found in the usual NLSE of cubic medium, such as negative sign of Kerr-term, Raman and self-steepening like terms of controllable sign and magnitude [14]. The NLSE of cubic nonlinear media is one of the most studied model in nonlinear optics, specifically due to the tremendous practical importance of its solitary wave or soliton solutions [7]. Optical solitons or solitary waves are robust localized pulses originating from the delicate balance between the dispersion effect and the nonlinear effect, manifesting through the so-called optical Kerr effect. Ultrashort optical soliton solutions have been found in many generalized NLSE model for cubic media beyond the SVEA [2,24]. In this work, our goal is to obtain exact few-cycle optical solitary wave solutions to a mathematical model considering the combined action of quadratic, cubic and quintic nonlinearities in a large phase-mismatched second harmonic (SHG) process. It is worthwhile to note that solitons in uniform cubic-quintic media within SVEA have been studied in many works [25-27]. In passing, it is important to note that modulation instability analysis of Moses-Wise model have been carried out recently and the study shows that it is possible to control the formation of ultrashort solitons in a cascaded-quadratic-cubic medium beyond the SVEA regime[28]. Also, it should be noted that due to advances in fabrication technology many materials such as $CdS_xSe_{1-x}$-doped glass are available which possesses a considerable level of fifth-order susceptibility ($\chi^{(5)}$)[27]. In fact, quite recently experiments show that materials such as chalcogenide glass exhibits not only third- and fifth order nonlinearities but even seventh-order nonlinearity [29].

The Moses-Wise model [14,23] considers the combined action of quadratic and cubic nonlinearities in a phase-mismatched SHG process. The effects of third-order dispersion, fourth-order dispersion and quintic nonlinearity can be incorporated into the model quite straightforwardly. The coupled propagation equation for the fundamental field (FF) and second-harmonic field (SHF), in the limit of large wave-vector mismatch, can be reduced to a single nonlinear evolution equation for the fundamental field:

$$i\frac{\partial u}{\partial \xi} - \frac{\alpha}{2}\frac{\partial^2 u}{\partial \tau^2} - i\delta_3 \frac{\partial^3 u}{\partial \tau^3} + \delta_4 \frac{\partial^4 u}{\partial \tau^4} + \beta |u|^2 u + \gamma |u|^4 u + i\beta s u^2 \frac{\partial u^*}{\partial \tau} + i\rho |u|^2 \frac{\partial u}{\partial \tau} = 0 \quad (1)$$

Here $\xi$ and $\tau$ are the normalized propagation distance and time respectively, $u$ is the normalized amplitude of the fundamental field. $\alpha = \pm 1$, refers to the sign of the GVD parameter. On the other hand, $\delta_3, \delta_4, \beta, \gamma$ and $s$ represents the normalized coefficients for third order dispersion (TOD), fourth order dispersion (FOD), cubic nonlinearity, quintic nonlinearity and self-steepening respectively. $\rho$ is a parameter arising from the mix of $\chi^{(2)}$ and $\chi^{(3)}$ effects, explicitly defined in Refs.[14,23,28]. In order to find the exact solitary wave solution of Eq. (1), we rewrite Eq. (1) in the following form:

$$iu_z - a_1 u_{\tau\tau} - ia_2 u_{\tau\tau\tau} + a_3 u_{\tau\tau\tau\tau} + a_4 |u|^2 u + ia_5 u^2 u_\tau^* + ia_6 |u|^2 u_\tau + a_7 |u|^4 u = 0 \quad (2)$$

Here, $z = \xi L_D$, $L_D$ is the so called dispersion length, while $a_1 = \alpha/2L_D$, $a_2 = \delta_3/L_D$, $a_3 = \delta_4/L_D$, $a_5 = \gamma/L_D$, $a_6 = \beta s/L_D$ and $a_7 = \rho/L_D$. We begin our analysis by assuming a solution of Eq.(2) in the form

$$u(z,\tau) = \phi(\chi)e^{i(m\chi - kz)}, \quad (3)$$

where $\chi = (\tau - vz)$ is the travelling coordinate and $\phi$ is a real function of $\chi$. Substituting Eq. (3) in Eq. (2) and separating the real and imaginary parts we obtain:

$$a_3\phi''' + (3ma_2 - 6m^2a_3 - a_1)\phi'' + (a_1m^2 - m^3a_2 + a_3m^4 + mv)\phi \\ + (a_4 - ma_6 + ma_5)\phi^3 + a_7\phi^5 = 0 \tag{4}$$

and

$$(4ma_3 - a_2)\phi''' + (a_5 + a_6)\phi^2\phi' + (3a_2m^2 - 4a_3m^3 - 2a_1m - v)\phi' = 0. \tag{5}$$

Integration of Eq. (5) leads to

$$\phi'' = a\phi + b\phi^3 \tag{6}$$

with

$$a = \frac{4m^3a_3 + 2ma_1 - 3a_2m^2 + v}{4ma_3 - a_2} \tag{7}$$

and

$$b = -\frac{a_5 + a_6}{3(4ma_3 - a_2)}. \tag{8}$$

Integration constant in Eq. (6) is taken to be zero with the assumption $\phi'' = 0$ for $\phi = 0$.
Integration of Eq. (6) leads to

$$\phi'^2 = a\phi^2 + \frac{b}{2}\phi^4 + c_2. \tag{9}$$

From Eq. (6) and Eq. (9) we get

$$\phi''' = 6b^2\phi^5 + 10ab\phi^3 + (6bc_2 + a^2)\phi. \tag{10}$$

Now, putting the expressions for $\phi'''$ and $\phi''$ from Eqs. (10) and (6) in Eq. (4) we obtain a polynomial equation in $\phi$. Equating the coefficients of different powers of $\phi$ to zero, we have

$$\phi^5: \quad 6b^2a_3 + a_7 = 0 \tag{11}$$
$$\phi^3: \quad 10aba_3 + 3mba_2 - 6m^2ba_3 - a_1b + a_4 + ma_5 - ma_6 = 0 \tag{12}$$
$$\phi: \quad a^2a_3 + 6bc_2a_3 + 3maa_2 - 6m^2aa_3 - aa_1 + m^2a_1 - m^3a_2 + a_3m^4 + k + mv = 0 \tag{13}$$

Solving the above set of equations, we obtain

$$b = \pm\sqrt{-\frac{a_7}{6a_3}}, \tag{14}$$

$$v = \frac{(4ma_3 - a_2)}{10ba_3}(6m^2ba_3 + a_1b - 3mba_2 - a_4 - ma_5 + ma_6) \\ + (3m^2a_2 - 4m^3a_3 - 2ma_1) \tag{15}$$

and

$$k = -a^2a_3 - 6bc_2a_3 - 3maa_2 + 6m^2aa_3 + aa_1 - a_1m^2 \\ + m^3a_2 - a_3m^4 - mv. \tag{16}$$

Comparing Eqs. (8) and (14), we have

$$m = \frac{1}{12a_3}[3a_2 \mp a_5 + a_6\sqrt{-\frac{6a_3}{a_7}}] \tag{17}$$

where '-' is for $b>0$ and '+' is for $b<0$.

Now, Eq. (9) is a well-known first-order ordinary differential equation which can be solved for dark and bright soliton solutions for different parametric conditions [30].

For $a>0, b<0$ and $c_2 = 0$, Eq. (9) can be solved for bright soliton solutions of the form

$$\phi = (-\frac{2a}{b})^{\frac{1}{2}} \text{sech}(\sqrt{a}\chi). \tag{18}$$

The complete solution for Eq. (2) reads

$$u(z,\tau) = (-\frac{2a}{b})^{\frac{1}{2}} \text{sech}(\sqrt{a}\chi) e^{i(m\chi - kz)}, \tag{19}$$

with $\chi = (\tau - vz)$, and $a, b, v, k$ and $m$ are defined in Eqs. (7), (14), (15), (16) and (17) respectively.

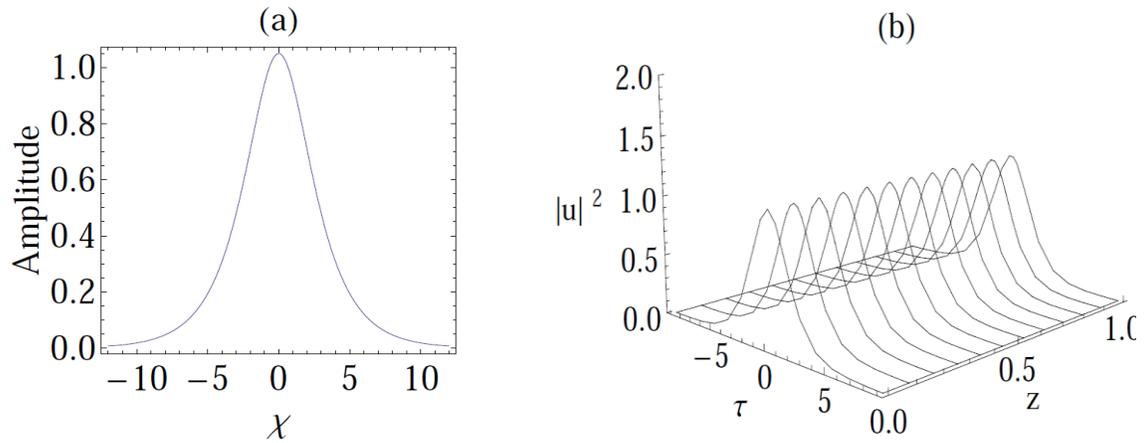

**Figure 1**. (a) Typical amplitude profile of bright soliton (b) Intensity evolution of bright soliton. The model parameters mentioned in the text.

The typical profiles for the amplitude and intensity of bright soliton are depicted in Figs. 1(a) and 1(b) respectively, for $a_1 = 1; a_2 = 2; a_3 = 1; a_4 = 1; a_5 = 2; a_6 = 1.2; a_7 = -1$. For these values, the wave parameters are found to be $a = 0.225$, $b = -0.408$, $m = 1.153$, $k = 0.236$ and $v = 0.126$. From plots, it is clear that the solitary wave propagates along the $z$-axis with same shape and amplitude, and it saturates at same finite value along time-axis as $\tau$ approaches its asymptotic value.

The analytical result is verified numerically by solving Eq.(2) using the well known split-step Fourier method. Fig. 2 depicts the temporal and spectral evolution of the bright soliton.

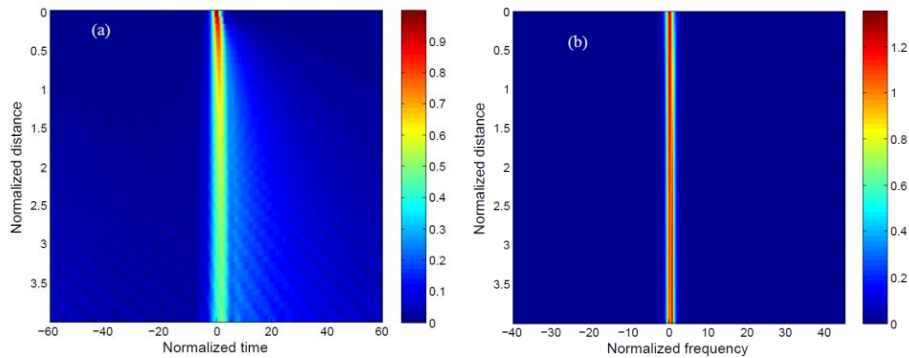

**Figure 2**. (a) Temporal and (b) spectral evolution of bright soliton in a cascaded quadratic-cubic-quintic media. Here we have taken the same set of parameters used in our analytical calculations. It is clear that numerical results agrees quite well with the analytical ones.

For $a<0, b>0$ and $c_2 = \dfrac{a^2}{2b}$, Eq. (9) can be solved for dark solitons given by

$$\phi = (-\dfrac{a}{b})^{\frac{1}{2}} \tanh[\sqrt{\dfrac{-a}{2}}\chi] \ . \tag{20}$$

For this case, the complete solution for Eq. (2) reads

$$u(z,\tau) = (-\dfrac{a}{b})^{\frac{1}{2}} \tanh[\sqrt{\dfrac{-a}{2}}\chi]e^{i(m\chi - kz)} \ . \tag{21}$$

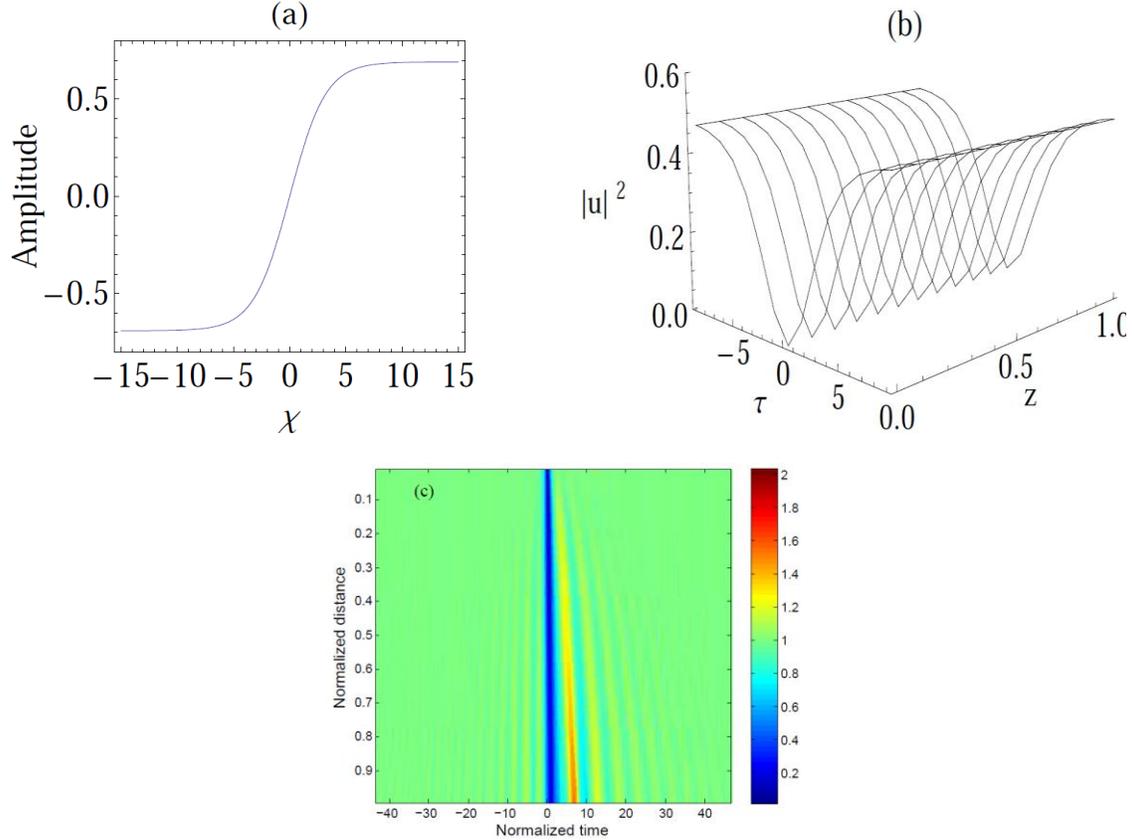

**Figure 3**. (a) Typical amplitude profile of dark soliton (b) Intensity evolution of dark soliton. The model parameters mentioned in the text (c) evolution of dark soliton: numerical simulation with the same model parameters as in (a) and (b)

The amplitude and intensity profiles for dark soliton are shown in Fig. 3(a) and 3(b) respectively, for $a_1 = 1; a_2 = 2; a_3 = 1; a_4 = 1; a_5 = 1; a_6 = 1; a_7 = -1$. For these values, the wave parameters are found to be $a$ = -0.195, $b$ = 0.408, $m$ = 0.092, $k$ = -0.159 and $v$ =0.182. The corresponding results obtained from direct numerical simulation of Eq.(2) is shown in Fig. 3(c).

      To conclude, we have studied the propagation of few-cycle optical solitary waves in a nonlinear media under the combined action of quadratic, cubic and quintic nonlinearities in a large phase-mismatched second harmonic (SHG) process. Exact bright and dark soliton solutions to the nonlinear evolution equation for cascaded quadratic media beyond the slowly varying

envelope approximations is reported. The analytical solutions obtained are found to be in good conformity with the numerical simulations.